\documentclass[twocolumn,superscriptaddress,amsmath,amssymb,aps,pra]{revtex4-2}

\usepackage{graphicx}
\usepackage{dcolumn}
\usepackage{bm}

\usepackage{graphicx}
\usepackage{setspace}
\usepackage{amsmath}
\usepackage{amssymb}
\usepackage{color}
\usepackage{array}
\usepackage{subfigure}
\usepackage{hyperref}
\usepackage{float}
\usepackage{lipsum}

\usepackage[all]{xy}
\newcommand{\RN}[1]{
\textup{\uppercase\expandafter{\romannumeral#1}}%
}

\begin{document}
\title{Quasi one-dimensional diffuse laser cooling of atoms}
\author{Jin-Yin Wan}
\affiliation{Laboratory of Space Laser Engineering and Technology, Shanghai Institute of Optics and Fine Mechanics, Chinese Academy of Sciences, Shanghai 201800, China}
\author{Xin Wang}
\affiliation{Key Laboratory of Quantum Optics and Center of Cold Atom Physics, Shanghai Institute of Optics and Fine Mechanics, Chinese Academy of Sciences, Shanghai 201800, China}
\author{Xiao Zhang}
\affiliation{Key Laboratory of Quantum Optics and Center of Cold Atom Physics, Shanghai Institute of Optics and Fine Mechanics, Chinese Academy of Sciences, Shanghai 201800, China}
\author{Yan-Ling Meng}
\affiliation{Laboratory of Space Laser Engineering and Technology, Shanghai Institute of Optics and Fine Mechanics, Chinese Academy of Sciences, Shanghai 201800, China}
\author{Wen-Li Wang}
\affiliation{Laboratory of Space Laser Engineering and Technology, Shanghai Institute of Optics and Fine Mechanics, Chinese Academy of Sciences, Shanghai 201800, China}
\author{Yuan Sun}
\email{sunyuan@siom.ac.cn}
\affiliation{Key Laboratory of Quantum Optics and Center of Cold Atom Physics, Shanghai Institute of Optics and Fine Mechanics, Chinese Academy of Sciences, Shanghai 201800, China}
\author{Liang Liu}
\email{liang.liu@siom.ac.cn}
\affiliation{Laboratory of Space Laser Engineering and Technology, Shanghai Institute of Optics and Fine Mechanics, Chinese Academy of Sciences, Shanghai 201800, China}
\affiliation{Key Laboratory of Quantum Optics and Center of Cold Atom Physics, Shanghai Institute of Optics and Fine Mechanics, Chinese Academy of Sciences, Shanghai 201800, China}

\begin{abstract}
We demonstrate experimentally the generation of one-dimensional cold gases of $^{87}$Rb atoms by diffuse laser cooling (DLC).  A horizontal slender vacuum glass tube with length of 105~cm and diameter of 2~cm is used in our experiment. The diffuse laser light inside the tube, which is generated by multi-reflection of injected lasers, cools the background vapor atoms.  With 250~mW  of cooling light and 50~mW of repumping light, an evenly distributed meter-long profile of atom cloud is obtained. We observe a factor 4 improvement on the atomic OD for a typical cooling duration of 170~ms and a sub-Doppler atomic temperature of 25~$\mu$k. The maximum number of detected cold atoms remain constant for a free-fall duration of 30~ms. Such samples are ideal for many quantum optical experiments involving electromagnetically induced transparency, electronically highly excited (Rydberg) atoms and quantum precision measurements.
\end{abstract}
\maketitle

\section{INTRODUCTION}\label{sec:intro}

Laser cooling techniques have revolutionized experimental research on cold atom physics and become a standard tool for precision measurement, quantum optics, and quantum simulation, quantum computation and quantum information~\cite{metcalf99book}. One typical laser cooling apparatus is the six-beam magneto-optical trap (MOT) requiring careful alignment of cooling laser beams and magnetic gradient field, which makes the development of precise metrological devices based on cold atoms such as atomic clocks, gravimeters and gyroscopes difficult and complex. Diffuse laser cooling (DLC) scheme is proposed to ease these problems, where the diffuse light field is generated in a tube or a sphere whose inner surface reflect laser light diffusely~\cite{ketterle1992slowing,hongxin1994laser,wang1995laser,batelaan1994slowing}.  Then the 3D diffuse light cooling configurations have been realised on Cs and Rb vapors in spherical cells~\cite{guillot2001three,cheng2009laser}. It shows a Sisyphus-like cooling mechanism since the cold atoms can reach a sub-Doppler temperature as low as 3.5~$\mu$k~\cite{guillot2001three,horak1998Atom,grynberg2000Spatial}. DLC technique is easy and robust since it needs no alignments of lasers and is not sensitive to magnetic field. DLC has been widely applied to compact microwave atomic clocks for ground operation and on-board applications in microgravity environment.~\cite{esnault2010high,ben2013Development,liu2015scheme,langlois2018Compact,wang2018optical}. Moreover, compact cold atom platform provides exciting possibilities for portable, robust and accessible quantum sensors. Therefore, neutral cold atom ensemble by DLC technique is also a potential candidate for space applications~\cite{van2010bose,liu2018orbit,aveline2020-Observation}.

DLC technique has been studied both in slowing an atomic beam and in cooling of atoms from vapor background ~\cite{ketterle1992slowing,hongxin1994laser,wang1995laser,batelaan1994slowing,guillot2001three,cheng2009laser}. Many efforts have been made on the properties of cold atoms captured by DLC including temperature, atomic density, localization, scale and spectroscopic effects ~\cite{zhang2009Observation,zhang2009Nonlinear,xu2012measurement,meng2014increasing,ZhengBC2014LargeScale,wan2015Non,wang2015Recoil,wang2020Nearly}. However, much less attention was paid to one-dimensional configuration. In this work, we demonstrate a meter-long diffuse laser cooling apparatus in which $^{87}$Rb atoms were cooled in a nearly one dimensional shape by DLC. Different from the conventional DLC setups ~\cite{guillot2001three,cheng2009laser,wang2020Nearly}, we employ a horizontal slender glass tube whose outside surface is covered by a reflective material. Atoms are cooled by diffuse light inside the tube homogeneously.

The paper is organized as follows: In Sec.~\ref{sec:experiment}, the experimental configuration for the DLC of $^{87}$Rb atoms in the horizontal slender glass tube and the corresponding time sequence are presented. In Sec.~\ref{sec:results}, as a key parameter to determine the performance of the captured cold atoms, the dependence of the atomic OD on a variety of experimental parameters, including the cooling light power, the repumping light power, the detuning of cooling and probe light are measured. The cooling time and temperature measurement of the cold atoms are also described. The atomic density distributions of the cold atoms in the longitudinal direction is detected through a pair of moving Helmholtz coils. Finally, the paper is concluded in Sec.~\ref{sec:conclusion}.

\section{EXPERIMENTAL CONFIGURATION}\label{sec:experiment}

\begin{figure}[htbp]
\centering
\begin{tabular}{l}
\includegraphics[trim = 0mm 0mm 0mm 0mm, clip, width=0.5\textwidth]{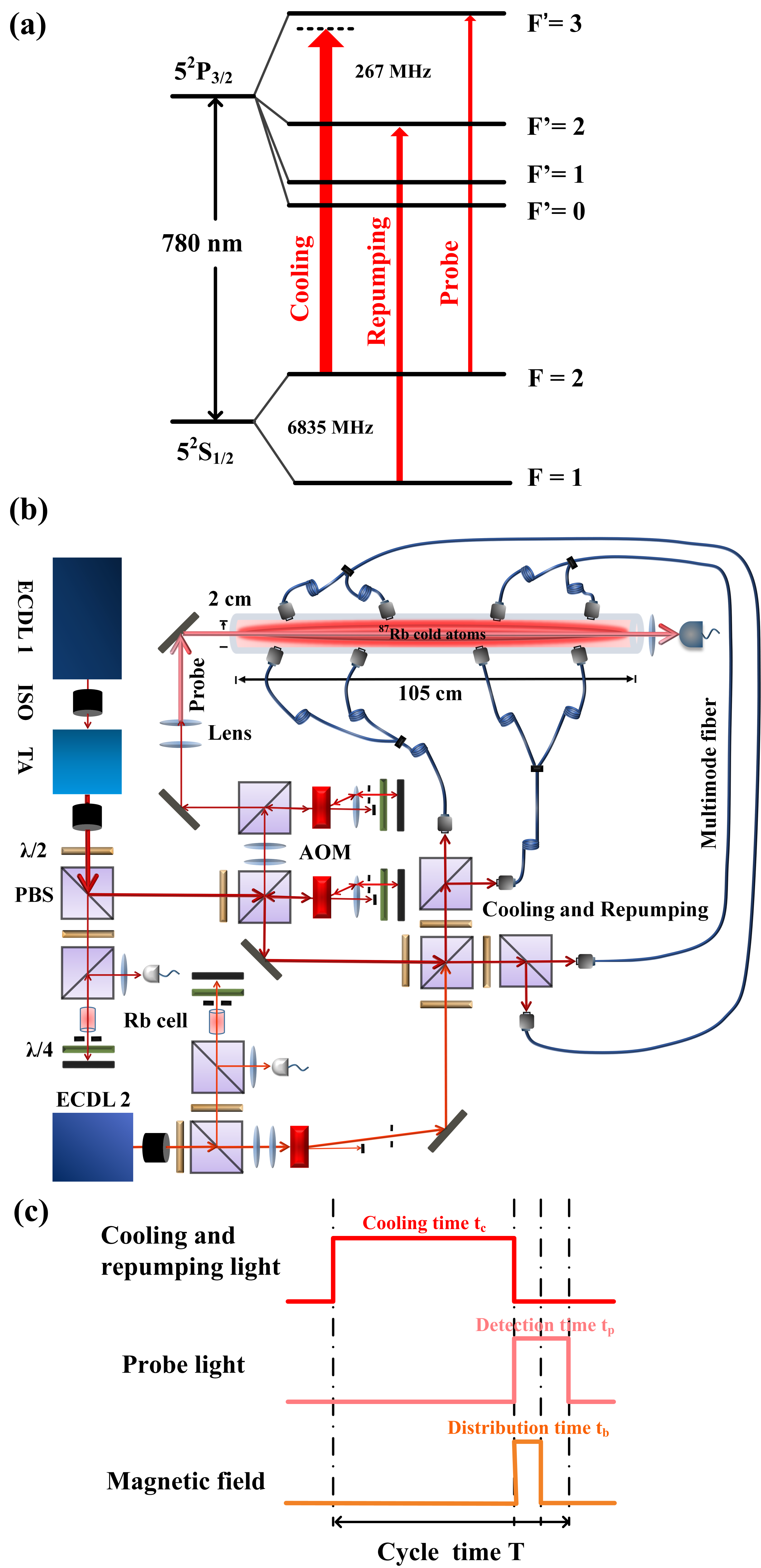}
\end{tabular}
\linespread{1}
\caption{DLC measurement scheme. (a) Relevant energy levels of $^{87}$Rb D$_{2}$ line for diffuse light cooling. (b) Schematic of the experimental setup. ECDL 1, extended-cavity diode laser 1, cooling and probe laser. ECDL 2, repumping laser. TA, tapered amplifier. ISO, isolator. PBS, polarization beam splitter. $\lambda/2$, half-wave plate. $\lambda/4$, quarter-wave plate. HR, high-reflection mirror. PD, photo detector. AOM, acoustic optical modulator. (c) DLC and atomic density distribution measurement timing sequence. In our case, the cooling time $t_{c}$ is 170~ms, the probe time $t_{p}$ is 100~ms, the distribution magnetic field time $t_{b}$ is 50~ms, and the cycle time T is 270~ms.
\label{physical_time}
}
\end{figure}

Figure~\ref{physical_time} gives the experimental scheme. Two extended-cavity diode lasers (ECDL) tuned at 780.24~nm of Rubidium D$_{2}$ line are used, both of which are frequency locked by saturation absorption spectra. One is used as both cooling and the probe, the other is used as the repumping, as shown in Fig.~\ref{physical_time} (a). Each laser frequency is individually controlled by acousto-optic modulators (AOMs).

Similar to traditional optical molasses, there are 4 energy levels involved of $^{87}$Rb atoms in our DLC experiment. The cooling light is red detuned by around -17~MHz ($\sim$2.8$\Gamma$, where $\Gamma$ = 6.056~MHz denotes the natural linewidth of $^{87}$Rb) to the transition $5^{2}S_{1/2}, F=2 \rightarrow 5^{2}P_{3/2}, F'=3$. The repumping light is resonant with the transition $5^{2}S_{1/2}, F=1 \rightarrow 5^{2}P_{3/2}, F'=2$ to prevent $^{87}$Rb atoms from accumulating in the $5^{2}S_{1/2},F=1$ energy level. The probe light is nearly resonant with the transition $5^{2}S_{1/2}, F=2 \rightarrow 5^{2}P_{3/2}, F'=3$.

\begin{figure}[thbp]
\centering
\begin{tabular}{l}
\includegraphics[trim = 0mm 0mm 0mm 0mm, clip, width=0.45\textwidth]{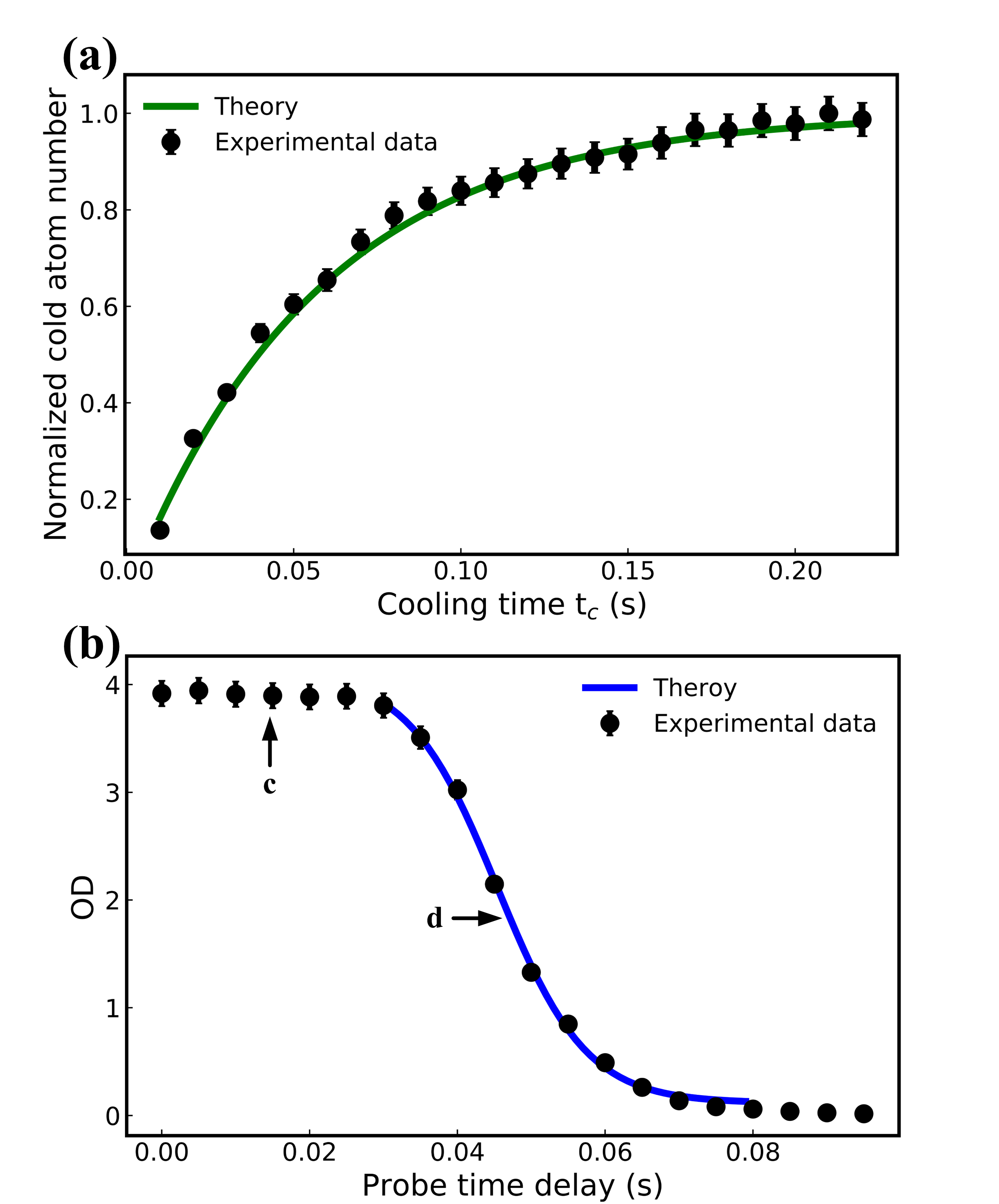}
\end{tabular}
\linespread{1}
\caption{(a) Normalized cold atom number along the optical detection axis versus the cooling time $t_{c}$, where the atom number has been normalized to a factor of 8 $\times$ 10$^{9}$. For cooling time more than about 0.17~s, saturation behavior occurs where the capturing and loss are almost balanced. (b) The measured OD at different probe time delays. A significant result of the maximum OD remains unchanged for a free-fall duration 30~ms, as indicated by the arrow (c). And arrow (d) indicates that it decreases exponentially with a theoretical fitting, where a temperature 25~$\mu$k is obtained.
\label{OD_time_probedelay}}
\end{figure}

A sketch of the experimental setup is shown in Fig. \ref{physical_time}(b), the main physical part is a horizontal cylindrical glass vacuum tube with its length of 105~cm and the inner cross section diameter of 2~cm, whose vacuum is maintained at $3\times10^{-8}$~Pa by an ion pump. Its outside surface is covered with a coating whose diffuse reflectance is more than 98\% at the optical wavelength of 780 nm. Sizable holes are scratched for the transmission of light beams. Homogeneous diffuse light inside the tube is formed from diffuse multi-reflection of lasers which are injected through 8 holes. The vacuum tube is filled with natural rubidium (Rb) vapor which is composed of $^{87}$Rb (27.8\%) and $^{85}$Rb (72.2\%). The Rb vapor is obtained by heating the Rb source to 70~$^{\texttt{o}}$C to release gaseous atoms. The source is then cooled down slowly to room temperature ($\sim$25$^{o}$C).

The optical system mainly consists of two extended-cavity diode lasers (ECDL), both of which are frequency locked by saturated absorption technique. One ECDL is locked at the transition $5^{2}S_{1/2}, F=2$ to the cross peak of $5^{2}P_{3/2}, F'=1$ and $5^{2}P_{3/2}, F'=3$. It is followed by a tapered amplifier (TA) which can emit the maximum light power of 2~W. The output beam is then divided into two beams via a polarizing beam splitter (PBS). One beam serves as the cooling light, and the other beam serves as the probe light. Each beam is switched and frequency shifted from a certain transition by using an acousto-optic modulator (AOM) in the double-pass configuration. The other ECDL is frequency locked at the transition $5^{2}S_{1/2}, F=1$ to the cross peak of $5^{2}P_{3/2}, F'=1$ and $5^{2}P_{3/2}, F'=2$. It serves as the repumping light which is controlled by an AOM in the single-pass configuration. Firstly, the cooling beam and repumping beam are combined by a PBS. Secondly, the two combined beams are divided to four balanced beams by another two PBSs. Thirdly, each beam is coupled into an One-to-Two multi-mode optical fiber, and eight intensity balanced divergent lights are generated. Finally, the fiber coupling heads are injected into the tube from the eight evenly distributed holes on the side surface, the injection angles can be between 30 and 60 degrees from the side section, thus there is no need to optimize the alignment of these beams. Compared with the well-designed optical alignments, this robust optical injection mode reduces the difficulty of the optical adjustment.

Fig.~\ref{physical_time}(c) shows the experimental sequence of the DLC process. The cooling and repumping lights are applied for 170~ms, this process will cool a large population of $^{87}$Rb atoms down from the background Rb vapor, the internal state of such $^{87}$Rb atoms is prepared to the energy level of $5^{2}S_{1/2}, F=2$. And then the probe light is applied for 100~ms to detect the captured cold atoms. In the first 50~ms of the probe time, a weak static magnetic field is applied to measure the atomic density distribution. The total cycle time is 270~ms.

\section{RESULTS AND DISCUSSIONS}\label{sec:results}

Atoms cooled by  DLC in the tube are detected by measuring absorption of a resonant probe laser in the direction of the tube's axis, as shown in Fig.~\ref{physical_time}(b). The The diameter of Gaussian probe beam is 1~mm with intensity around 100~$\mu$W/cm$^{2}$. Generally, OD of the cold atoms is given by the relation $I=I_{0}e^{-OD}$, here $I_{0}$ and $I$ are the incident and the output intensity, respectively. The number of  cold atoms $N$ obeys the relation $OD=\sigma N$, where $\sigma$ is the cross section of the atom-light interaction. The maximum number of cold atoms can be reached at the balance between capture and loss of laser-cooled atoms . A straightforward phenomenological model may be given as $dN/dt = -\kappa N + \alpha$, where $\kappa$ describes the loss and $\alpha$ describes the replenishing rate of the newly captured cold atoms~\cite{metcalf99book}. Therefore $N(t) = \alpha/\kappa (1-e^{-\kappa (t-t_{0})})$, and $N(\infty) = \alpha / \kappa$ when $t$ is long enough.

Fig.~\ref{OD_time_probedelay}(a) shows the normalized cold atom number versus the cooling time $t_{c}$. Simulation, shown in the solid line in Fig..~\ref{OD_time_probedelay}(a) ,  gives the loss rate $\kappa$  as 17.6 s$^{-1}$. When the cooling time lasts more than 0.17~s, the number of  cold atoms reaches saturation, which is about 6 times less than the previous typical DLC schemes ~\cite{wang2020Nearly}. The fast banance comes from the quick loss of cooled atoms due to the small diameter of the tube which is horizontally placed in the gravity. 

In order to determine the temperature of the captured cold atoms, we use a recently developed nearly nondestructive thermometry of labeled cold atoms. In this work, we roughly regard that all the detected atoms after the cooling process are at the energy level of $5^{2}S_{1/2}, F=2$, neglecting the polarization degree of freedoms. The number of cold atoms within the beam size of the probe light will decrease over probe time caused by spatial expansion and gravity. The effect of gravity is neglected for simplicity since the expansion time of the labeled atoms is rather short. The temperature of the cold atom ensemble can be evaluated by the detection of the probe light over time. Moreover, dynamics along the axis of the slender cavity rarely affect the signal. Therefore the process can be considered under a 2D setting in the transverse plane. And the velocity distribution of the cold atoms is well described by the Maxwell-Boltzmann
distribution $f(v)dv = (\frac{m}{2\pi \kappa_{B} T})2\pi v \exp (- \frac{m v^{2}}{2\kappa_{B} T}) dv$, the cold atom number detected in the range of the probe beam with radius of $r$ is proportional to~\cite{wang2020Nearly}:

\begin{equation}\label{no_t}
  s(r,t) = 1-\exp(-\frac{m(r/t)^{2}}{2k_{B}T}).
\end{equation}
\par
The experimental results and fitting are shown in Fig. \ref{OD_time_probedelay}(b). Arrow (c) indicates that the measured OD almost keeps at the maximum for the free-fall duration of 30~ms, since the density distribution of the targeted cold atoms on the optical detection axis remain unchanged. It is the first observation of a nearly fixed maximum cold atom number in a tens of milliseconds time range once the cooling light is switched off. Generally, the captured cold atoms starts to decrease immediately when there is no cooling light~\cite{cheng2009laser}, which would be limited by the gravity and thermal expansion of the cloud. Arrow (d) indicates the thermal expansion process of addressed atoms with the theoretical fitting curve according to Eq. (\ref{no_t}), where the temperature is deduced to be 25~$\mu$k. It demonstrates that a sub-Doppler cooling like mechanism exists which is caused by the disordered gradient forces formed in the diffuse light field.

\begin{figure}[thbp]
\centering
\begin{tabular}{l}
\includegraphics[trim = 0mm 0mm 0mm 0mm, clip, width=0.5\textwidth]{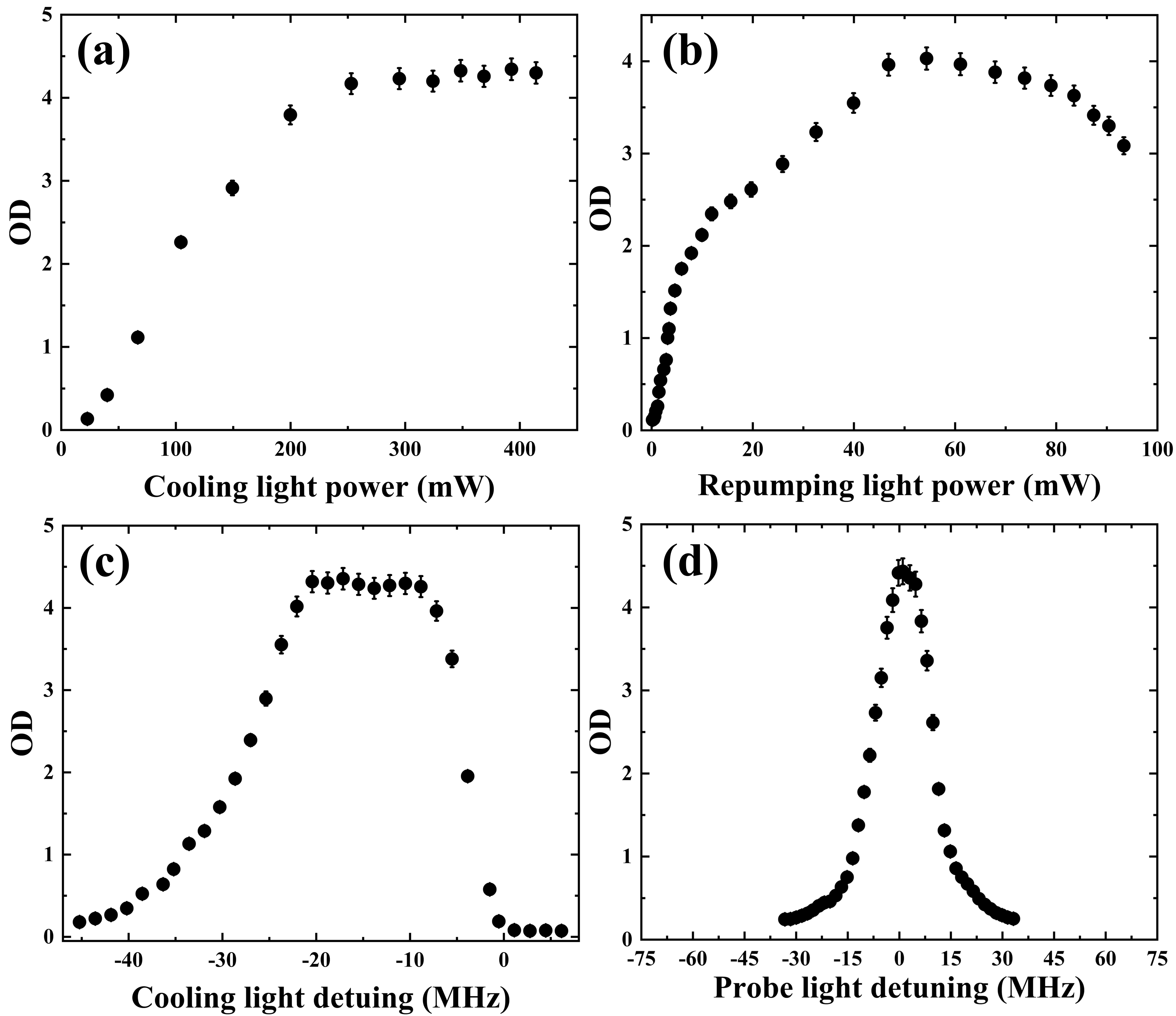}
\end{tabular}
\linespread{1}
\caption{The measured OD as a function of various cooling parameters. (a) OD versus cooling light power when  the cooling light is red detuned at 12~MHz and the repumpming light power is fixed at 50~mW.  (b) OD versus repumping light power when the when the cooling light is red detuned at 12~MHz and the cooling light power is fixed at 250~mW. (c) OD versus cooling light detuning when the the cooling light power is 250~mW and the repumpming light power is fixed at 50~mW. (d) OD versus probe light detuning when the cooling light is red detuned at 17~MHz, the the cooling light power is 250~mW and the repumpming light power is fixed at 50~mW.
\label{od_all}}
\end{figure}

We measured  the OD vs the total power of lasers injected into the vacuum tube with fixed detuning of -12~MHz, cooling time $t_{c}$ is 0.17~s, the probe time is 0.1~s, and the repumping power at 50~mW. As shown in Fig. \ref{od_all}(a), the OD increases linearly with the cooling light power, and it reaches a saturation value as the cooling light power is more than 250~mW.

Then we set the cooling light power at 250~mW and vary the repumping light power. Fig. \ref{od_all}(b) indicates that the OD increases with the repumping light power in the lower power region, and it reaches saturation at the value of 50~mW. However, a slight decrease is monitored in the higher power region, which is
caused by the heating effect. As a result, we obtain the atomic OD greater than 4 with the cooling light power of only 250~mW, such low laser power reduces the complicity of the laser system.

Fig.~\ref{od_all}(c) shows the measured OD as a function of cooling laser detuning. The measured OD almost maintains at a maximum value of 4.3for the detuning rangeing between -20 and -8.8~MHz (about -3.3$\Gamma$ to -1.4$\Gamma$). Different from a MOT or optical molasses, the velocity capture range of DLC is $\mid\Delta\mid\leq kv\leq2\mid\Delta\mid$~\cite{ketterle1992slowing}, where $\Delta$ denotes the detuning of the cooling light, $k$ is the wave vector, and $v$ is the atomic velocity. Such a wide capture range allows capturing atoms with large range of velocities. 

We measured the atomic OD versus the probe light detuning as shown in Fig. \ref{od_all}(d). The measured OD reaches the largest value of 4.4 when the probe light is resonant with the atomic transition, and therefore all measurements keep the probe light resonant for the estimation of number of cold atoms.

\begin{figure}[thbp]
 \centering
\begin{tabular}{l}
\includegraphics[trim = 0mm 0mm 0mm 0mm, clip, width=0.5\textwidth]{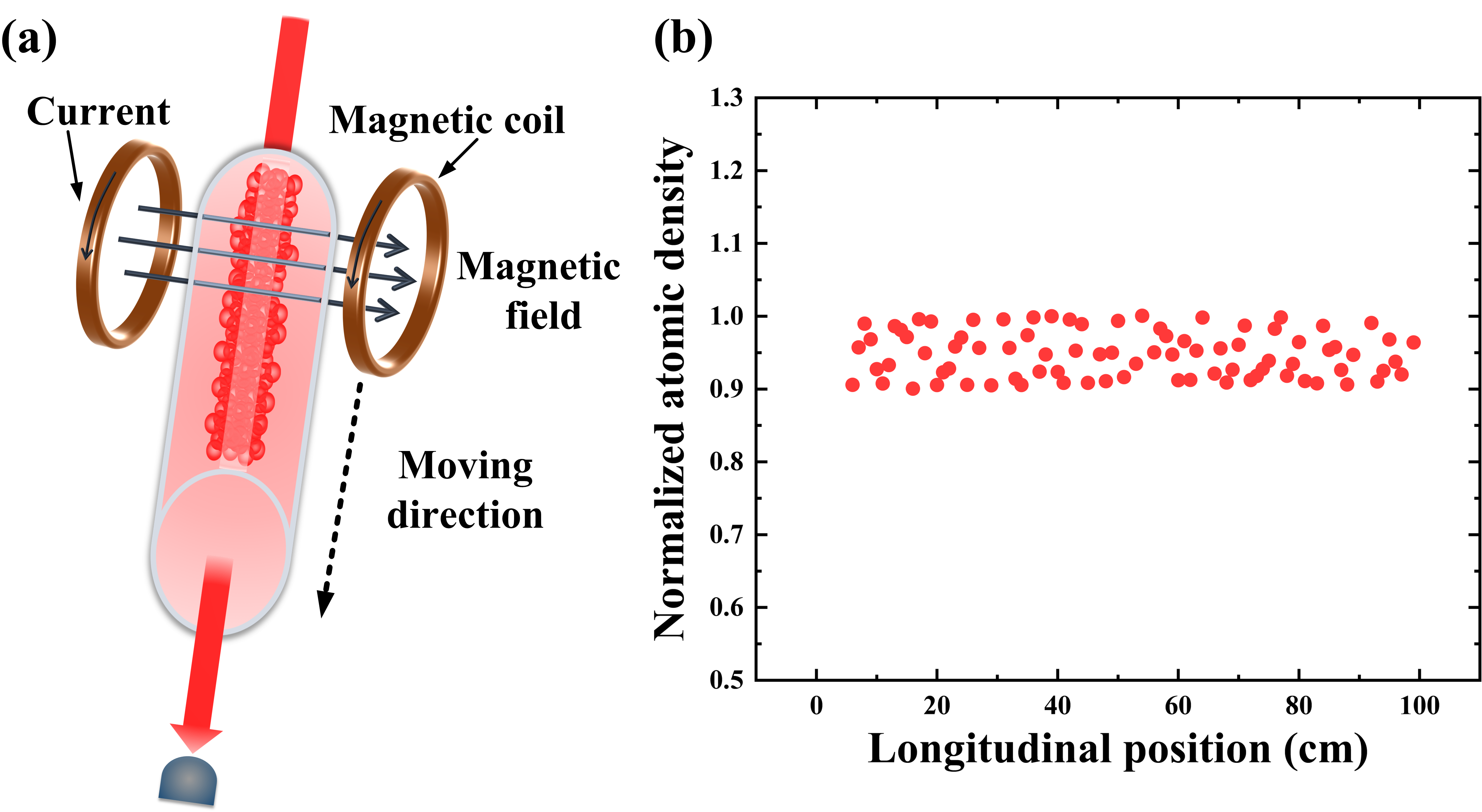}
\end{tabular}
\linespread{1}
\caption{The atomic density measurement scheme. (a) The atomic density measurement setup. A pair of Helmholtz coils is used to measure the longitudinal atomic density distribution. (b) The measured atomic density distribution along the longitudinal direction of the cavity.
\label{distribution}}
\end{figure}

To measure the atomic density distribution along the probe direction, we apply a weak uniform static magnetic field ($\sim$2.3~mT) perpendicular to the tube's longitudinal axis as shown in Fig.~\ref{distribution}(a). The magnetic field is generated by a pair of Helmholtz coils which are distributed on both sides of the tube perpendicularly and equidistantly. Generally, the resonant probe light measures the atoms at the $5^{2}S_{1/2}, F=2$ energy level. When the atoms' energy level is shifted from resonance by 30~MHz through a static magnetic field as shown in Fig.~\ref{od_all}(d), the part of cold atoms in the magnetic field region would not be detected by the probe light. The data of these cold atoms can be directly obtained when it is subtracted by the total signal detected without magnetic field, the data of cold atoms in the magnetic field region can be obtained. Therefore, we measured the cold atom density distribution by moving the helmholtz coils from one end of the tube to the other. In the first 30~ms after the cooling light is switched off, the atomic distribution signal is too small to be detected precisely since the cold atom number keeps the maximum and stable in this time region as shown in Fig. \ref{OD_time_probedelay}(b). Thus the atomic density distribution is measured at the probe time delay of 40~ms as shown in Fig.~\ref{distribution}(b). The cold atom number distribution lies in a horizontal line along the longitudinal direction of the tube, thus a meter-long nearly one dimensional cold atom cloud is prepared. Since the tube is slender, the light emitted by the multi-mode optical fibers travels very short distance before touching the wall and then is diffusely reflected at a high rate immediately. The isotropic lights in the horizontal chamber forms a quasi one-dimensional field which is more evenly distributed than that in the spherical or vertical configurations~\cite{meng2014increasing}, and it homogenizes the atomic density.

\section{CONCLUSION}\label{sec:conclusion}
In conclusion, we have experimentally demostrated nearly one dimensional cold $^{87}$Rb atom cloud with a length of 105~cm and a diameter of 2~cm by DLC in a horizontally-placed slender glass tube.   We have systematically studied the dependence of atomic OD on several control parameters. With a total cooling power of 250 mW, repumping power of 50 mW and a cooling time of 0.17~s, we obtain a horizontally uniform distributed cold $^{87}$Rb atom ensemble with the atomic OD more than 4 and the temperature about 25~$\mu$k. We first observe that the maximum atomic OD almost remain unchanged for a free-fall duration 30~ms.

We have also shown the simplicity and robustness of the DLC. In fact, the DLC depends only on the interaction between diffuse light and vapor atoms, regardless of scale, shape, polarization, propagation direction of laser field, magnetic field and etc. Such characters make the DLC easy to construct different kind of geometries of cold atom cloud~\cite{ZhengBC2014LargeScale} which are not possible by MOT or optical molasses, for example, the cold atom cloud in meter-long tube in this work can be easily extended to km-long or even longer, and also to 2D shape.

Our results give a new method to generate cold atoms especially in some geometries, which are useful in quantum sensing.

\par

\begin{acknowledgments}
We thank Weibin Li for helpful suggestions in the preparation of the manuscript. This work is supported by the National Natural Science Foundation of China (Grant No. 61727821).
\end{acknowledgments}

\bibliography{DiffuseCoolingRef}

\begin{thebibliography}{25}%
\makeatletter
\providecommand \@ifxundefined [1]{%
 \@ifx{#1\undefined}
}%
\providecommand \@ifnum [1]{%
 \ifnum #1\expandafter \@firstoftwo
 \else \expandafter \@secondoftwo
 \fi
}%
\providecommand \@ifx [1]{%
 \ifx #1\expandafter \@firstoftwo
 \else \expandafter \@secondoftwo
 \fi
}%
\providecommand \natexlab [1]{#1}%
\providecommand \enquote  [1]{``#1''}%
\providecommand \bibnamefont  [1]{#1}%
\providecommand \bibfnamefont [1]{#1}%
\providecommand \citenamefont [1]{#1}%
\providecommand \href@noop [0]{\@secondoftwo}%
\providecommand \href [0]{\begingroup \@sanitize@url \@href}%
\providecommand \@href[1]{\@@startlink{#1}\@@href}%
\providecommand \@@href[1]{\endgroup#1\@@endlink}%
\providecommand \@sanitize@url [0]{\catcode `\\12\catcode `\$12\catcode
  `\&12\catcode `\#12\catcode `\^12\catcode `\_12\catcode `\%12\relax}%
\providecommand \@@startlink[1]{}%
\providecommand \@@endlink[0]{}%
\providecommand \url  [0]{\begingroup\@sanitize@url \@url }%
\providecommand \@url [1]{\endgroup\@href {#1}{\urlprefix }}%
\providecommand \urlprefix  [0]{URL }%
\providecommand \Eprint [0]{\href }%
\providecommand \doibase [0]{https://doi.org/}%
\providecommand \selectlanguage [0]{\@gobble}%
\providecommand \bibinfo  [0]{\@secondoftwo}%
\providecommand \bibfield  [0]{\@secondoftwo}%
\providecommand \translation [1]{[#1]}%
\providecommand \BibitemOpen [0]{}%
\providecommand \bibitemStop [0]{}%
\providecommand \bibitemNoStop [0]{.\EOS\space}%
\providecommand \EOS [0]{\spacefactor3000\relax}%
\providecommand \BibitemShut  [1]{\csname bibitem#1\endcsname}%
\let\auto@bib@innerbib\@empty
\bibitem [{\citenamefont {Metcalf}\ and\ \citenamefont {Van~der
  Straten}(1999)}]{metcalf99book}%
  \BibitemOpen
  \bibfield  {author} {\bibinfo {author} {\bibfnamefont {H.~J.}\ \bibnamefont
  {Metcalf}}\ and\ \bibinfo {author} {\bibfnamefont {P.}~\bibnamefont {Van~der
  Straten}},\ }\href@noop {} {\emph {\bibinfo {title} {Laser Cooling and
  Trapping}}}\ (\bibinfo  {publisher} {Springer-Verlag New York, Inc.},\
  \bibinfo {year} {1999})\BibitemShut {NoStop}%
\bibitem [{\citenamefont {Ketterle}\ \emph {et~al.}(1992)\citenamefont
  {Ketterle}, \citenamefont {Martin}, \citenamefont {Joffe},\ and\
  \citenamefont {Pritchard}}]{ketterle1992slowing}%
  \BibitemOpen
  \bibfield  {author} {\bibinfo {author} {\bibfnamefont {W.}~\bibnamefont
  {Ketterle}}, \bibinfo {author} {\bibfnamefont {A.}~\bibnamefont {Martin}},
  \bibinfo {author} {\bibfnamefont {M.~A.}\ \bibnamefont {Joffe}},\ and\
  \bibinfo {author} {\bibfnamefont {D.~E.}\ \bibnamefont {Pritchard}},\
  }\bibfield  {title} {\bibinfo {title} {Slowing and cooling atoms in isotropic
  laser light},\ }\href@noop {} {\bibfield  {journal} {\bibinfo  {journal}
  {Phy. Rev. Lett.}\ }\textbf {\bibinfo {volume} {69}},\ \bibinfo {pages}
  {2483} (\bibinfo {year} {1992})}\BibitemShut {NoStop}%
\bibitem [{\citenamefont {Chen}\ \emph {et~al.}(1994)\citenamefont {Chen},
  \citenamefont {Cai}, \citenamefont {Liu}, \citenamefont {Shu}, \citenamefont
  {Li},\ and\ \citenamefont {Wang}}]{hongxin1994laser}%
  \BibitemOpen
  \bibfield  {author} {\bibinfo {author} {\bibfnamefont {H.-X.}\ \bibnamefont
  {Chen}}, \bibinfo {author} {\bibfnamefont {W.-Q.}\ \bibnamefont {Cai}},
  \bibinfo {author} {\bibfnamefont {L.}~\bibnamefont {Liu}}, \bibinfo {author}
  {\bibfnamefont {W.}~\bibnamefont {Shu}}, \bibinfo {author} {\bibfnamefont
  {F.-S.}\ \bibnamefont {Li}},\ and\ \bibinfo {author} {\bibfnamefont {Y.-Z.}\
  \bibnamefont {Wang}},\ }\bibfield  {title} {\bibinfo {title} {Laser cooling
  and deceleration of neutral atoms by red-shifted diffuse light},\ }\href@noop
  {} {\bibfield  {journal} {\bibinfo  {journal} {Chin. Phys. Lett.}\ }\textbf
  {\bibinfo {volume} {11}},\ \bibinfo {pages} {541} (\bibinfo {year}
  {1994})}\BibitemShut {NoStop}%
\bibitem [{\citenamefont {Wang}\ and\ \citenamefont
  {Liu}(1995)}]{wang1995laser}%
  \BibitemOpen
  \bibfield  {author} {\bibinfo {author} {\bibfnamefont {Y.-Z.}\ \bibnamefont
  {Wang}}\ and\ \bibinfo {author} {\bibfnamefont {L.}~\bibnamefont {Liu}},\
  }\bibfield  {title} {\bibinfo {title} {Laser manipulation of atoms and atom
  optics},\ }\href@noop {} {\bibfield  {journal} {\bibinfo  {journal} {Aust. J.
  Phys.}\ }\textbf {\bibinfo {volume} {48}},\ \bibinfo {pages} {267} (\bibinfo
  {year} {1995})}\BibitemShut {NoStop}%
\bibitem [{\citenamefont {Batelaan}\ \emph {et~al.}(1994)\citenamefont
  {Batelaan}, \citenamefont {Padua}, \citenamefont {Yang}, \citenamefont {Xie},
  \citenamefont {Gupta},\ and\ \citenamefont {Metcalf}}]{batelaan1994slowing}%
  \BibitemOpen
  \bibfield  {author} {\bibinfo {author} {\bibfnamefont {H.}~\bibnamefont
  {Batelaan}}, \bibinfo {author} {\bibfnamefont {S.}~\bibnamefont {Padua}},
  \bibinfo {author} {\bibfnamefont {D.}~\bibnamefont {Yang}}, \bibinfo {author}
  {\bibfnamefont {C.}~\bibnamefont {Xie}}, \bibinfo {author} {\bibfnamefont
  {R.}~\bibnamefont {Gupta}},\ and\ \bibinfo {author} {\bibfnamefont
  {H.}~\bibnamefont {Metcalf}},\ }\bibfield  {title} {\bibinfo {title} {Slowing
  of rb 85 atoms with isotropic light},\ }\href@noop {} {\bibfield  {journal}
  {\bibinfo  {journal} {Phy. Rev. A}\ }\textbf {\bibinfo {volume} {49}},\
  \bibinfo {pages} {2780} (\bibinfo {year} {1994})}\BibitemShut {NoStop}%
\bibitem [{\citenamefont {Guillot}\ \emph {et~al.}(2001)\citenamefont
  {Guillot}, \citenamefont {Pottie},\ and\ \citenamefont
  {Dimarcq}}]{guillot2001three}%
  \BibitemOpen
  \bibfield  {author} {\bibinfo {author} {\bibfnamefont {E.}~\bibnamefont
  {Guillot}}, \bibinfo {author} {\bibfnamefont {P.-E.}\ \bibnamefont
  {Pottie}},\ and\ \bibinfo {author} {\bibfnamefont {N.}~\bibnamefont
  {Dimarcq}},\ }\bibfield  {title} {\bibinfo {title} {Three-dimensional cooling
  of cesium atoms in a reflecting copper cylinder},\ }\href@noop {} {\bibfield
  {journal} {\bibinfo  {journal} {Opt. Lett.}\ }\textbf {\bibinfo {volume}
  {26}},\ \bibinfo {pages} {1639} (\bibinfo {year} {2001})}\BibitemShut
  {NoStop}%
\bibitem [{\citenamefont {Cheng}\ \emph {et~al.}(2009)\citenamefont {Cheng},
  \citenamefont {Zhang}, \citenamefont {Ma}, \citenamefont {Liu},\ and\
  \citenamefont {Wang}}]{cheng2009laser}%
  \BibitemOpen
  \bibfield  {author} {\bibinfo {author} {\bibfnamefont {H.-D.}\ \bibnamefont
  {Cheng}}, \bibinfo {author} {\bibfnamefont {W.-Z.}\ \bibnamefont {Zhang}},
  \bibinfo {author} {\bibfnamefont {H.-Y.}\ \bibnamefont {Ma}}, \bibinfo
  {author} {\bibfnamefont {L.}~\bibnamefont {Liu}},\ and\ \bibinfo {author}
  {\bibfnamefont {Y.-Z.}\ \bibnamefont {Wang}},\ }\bibfield  {title} {\bibinfo
  {title} {Laser cooling of rubidium atoms from background vapor in diffuse
  light},\ }\href@noop {} {\bibfield  {journal} {\bibinfo  {journal} {Phys.
  Rev. A}\ }\textbf {\bibinfo {volume} {79}},\ \bibinfo {pages} {023407}
  (\bibinfo {year} {2009})}\BibitemShut {NoStop}%
\bibitem [{\citenamefont {Horak}\ \emph {et~al.}(1998)\citenamefont {Horak},
  \citenamefont {Courtois},\ and\ \citenamefont {Grynberg}}]{horak1998Atom}%
  \BibitemOpen
  \bibfield  {author} {\bibinfo {author} {\bibfnamefont {P.}~\bibnamefont
  {Horak}}, \bibinfo {author} {\bibfnamefont {J.-Y.}\ \bibnamefont
  {Courtois}},\ and\ \bibinfo {author} {\bibfnamefont {G.}~\bibnamefont
  {Grynberg}},\ }\bibfield  {title} {\bibinfo {title} {Atom cooling and
  trapping by disorder},\ }\href@noop {} {\bibfield  {journal} {\bibinfo
  {journal} {Phys. Rev. A}\ }\textbf {\bibinfo {volume} {58}},\ \bibinfo
  {pages} {3953} (\bibinfo {year} {1998})}\BibitemShut {NoStop}%
\bibitem [{\citenamefont {Grynberg}\ \emph {et~al.}(2000)\citenamefont
  {Grynberg}, \citenamefont {Horak},\ and\ \citenamefont
  {Mennerat-Robilliard}}]{grynberg2000Spatial}%
  \BibitemOpen
  \bibfield  {author} {\bibinfo {author} {\bibfnamefont {G.}~\bibnamefont
  {Grynberg}}, \bibinfo {author} {\bibfnamefont {P.}~\bibnamefont {Horak}},\
  and\ \bibinfo {author} {\bibfnamefont {C.}~\bibnamefont
  {Mennerat-Robilliard}},\ }\bibfield  {title} {\bibinfo {title} {Spatial
  diffusion of atoms cooled in a speckle field},\ }\href@noop {} {\bibfield
  {journal} {\bibinfo  {journal} {Europhys. Lett.}\ }\textbf {\bibinfo {volume}
  {49}},\ \bibinfo {pages} {424} (\bibinfo {year} {2000})}\BibitemShut
  {NoStop}%
\bibitem [{\citenamefont {Esnault}\ \emph {et~al.}(2010)\citenamefont
  {Esnault}, \citenamefont {Holleville}, \citenamefont {Rossetto},
  \citenamefont {Guerandel},\ and\ \citenamefont {Dimarcq}}]{esnault2010high}%
  \BibitemOpen
  \bibfield  {author} {\bibinfo {author} {\bibfnamefont {F.-X.}\ \bibnamefont
  {Esnault}}, \bibinfo {author} {\bibfnamefont {D.}~\bibnamefont {Holleville}},
  \bibinfo {author} {\bibfnamefont {N.}~\bibnamefont {Rossetto}}, \bibinfo
  {author} {\bibfnamefont {S.}~\bibnamefont {Guerandel}},\ and\ \bibinfo
  {author} {\bibfnamefont {N.}~\bibnamefont {Dimarcq}},\ }\bibfield  {title}
  {\bibinfo {title} {High-stability compact atomic clock based on isotropic
  laser cooling},\ }\href@noop {} {\bibfield  {journal} {\bibinfo  {journal}
  {Phys. Rev. A}\ }\textbf {\bibinfo {volume} {82}},\ \bibinfo {pages} {033436}
  (\bibinfo {year} {2010})}\BibitemShut {NoStop}%
\bibitem [{\citenamefont {Zheng}\ \emph {et~al.}(2013)\citenamefont {Zheng},
  \citenamefont {Cheng}, \citenamefont {Meng}, \citenamefont {Xiao},
  \citenamefont {Wan},\ and\ \citenamefont {Liu}}]{ben2013Development}%
  \BibitemOpen
  \bibfield  {author} {\bibinfo {author} {\bibfnamefont {B.-C.}\ \bibnamefont
  {Zheng}}, \bibinfo {author} {\bibfnamefont {H.-D.}\ \bibnamefont {Cheng}},
  \bibinfo {author} {\bibfnamefont {Y.-L.}\ \bibnamefont {Meng}}, \bibinfo
  {author} {\bibfnamefont {L.}~\bibnamefont {Xiao}}, \bibinfo {author}
  {\bibfnamefont {J.-Y.}\ \bibnamefont {Wan}},\ and\ \bibinfo {author}
  {\bibfnamefont {L.}~\bibnamefont {Liu}},\ }\bibfield  {title} {\bibinfo
  {title} {Development of an integrating sphere cold atom clock},\ }\href@noop
  {} {\bibfield  {journal} {\bibinfo  {journal} {Chin. Phys. Lett.}\ }\textbf
  {\bibinfo {volume} {30}},\ \bibinfo {pages} {123701} (\bibinfo {year}
  {2013})}\BibitemShut {NoStop}%
\bibitem [{\citenamefont {Liu}\ \emph {et~al.}(2015)\citenamefont {Liu},
  \citenamefont {Meng}, \citenamefont {Wan}, \citenamefont {Wang},
  \citenamefont {Wang}, \citenamefont {Xiao}, \citenamefont {Cheng},\ and\
  \citenamefont {Liu}}]{liu2015scheme}%
  \BibitemOpen
  \bibfield  {author} {\bibinfo {author} {\bibfnamefont {P.}~\bibnamefont
  {Liu}}, \bibinfo {author} {\bibfnamefont {Y.-L.}\ \bibnamefont {Meng}},
  \bibinfo {author} {\bibfnamefont {J.-Y.}\ \bibnamefont {Wan}}, \bibinfo
  {author} {\bibfnamefont {X.-M.}\ \bibnamefont {Wang}}, \bibinfo {author}
  {\bibfnamefont {Y.-N.}\ \bibnamefont {Wang}}, \bibinfo {author}
  {\bibfnamefont {L.}~\bibnamefont {Xiao}}, \bibinfo {author} {\bibfnamefont
  {H.-D.}\ \bibnamefont {Cheng}},\ and\ \bibinfo {author} {\bibfnamefont
  {L.}~\bibnamefont {Liu}},\ }\bibfield  {title} {\bibinfo {title} {Scheme for
  a compact cold-atom clock based on diffuse laser cooling in a cylindrical
  cavity},\ }\href@noop {} {\bibfield  {journal} {\bibinfo  {journal} {Phy.
  Rev. A}\ }\textbf {\bibinfo {volume} {92}},\ \bibinfo {pages} {062101}
  (\bibinfo {year} {2015})}\BibitemShut {NoStop}%
\bibitem [{\citenamefont {Langlois}\ \emph {et~al.}(2018)\citenamefont
  {Langlois}, \citenamefont {De~Sarlo}, \citenamefont {Holleville},
  \citenamefont {Dimarcq}, \citenamefont {Schaff},\ and\ \citenamefont
  {Bernon}}]{langlois2018Compact}%
  \BibitemOpen
  \bibfield  {author} {\bibinfo {author} {\bibfnamefont {M.}~\bibnamefont
  {Langlois}}, \bibinfo {author} {\bibfnamefont {L.}~\bibnamefont {De~Sarlo}},
  \bibinfo {author} {\bibfnamefont {D.}~\bibnamefont {Holleville}}, \bibinfo
  {author} {\bibfnamefont {N.}~\bibnamefont {Dimarcq}}, \bibinfo {author}
  {\bibfnamefont {J.-F.}\ \bibnamefont {Schaff}},\ and\ \bibinfo {author}
  {\bibfnamefont {S.}~\bibnamefont {Bernon}},\ }\bibfield  {title} {\bibinfo
  {title} {Compact cold-atom clock for onboard timebase: Tests in reduced
  gravity},\ }\href@noop {} {\bibfield  {journal} {\bibinfo  {journal} {Phys.
  Rev. Applied}\ }\textbf {\bibinfo {volume} {10}},\ \bibinfo {pages} {064007}
  (\bibinfo {year} {2018})}\BibitemShut {NoStop}%
\bibitem [{\citenamefont {Wang}\ \emph {et~al.}(2018)\citenamefont {Wang},
  \citenamefont {Meng}, \citenamefont {Wan}, \citenamefont {Yu}, \citenamefont
  {Wang}, \citenamefont {Xiao}, \citenamefont {Cheng},\ and\ \citenamefont
  {Liu}}]{wang2018optical}%
  \BibitemOpen
  \bibfield  {author} {\bibinfo {author} {\bibfnamefont {Y.-N.}\ \bibnamefont
  {Wang}}, \bibinfo {author} {\bibfnamefont {Y.-L.}\ \bibnamefont {Meng}},
  \bibinfo {author} {\bibfnamefont {J.-Y.}\ \bibnamefont {Wan}}, \bibinfo
  {author} {\bibfnamefont {M.-Y.}\ \bibnamefont {Yu}}, \bibinfo {author}
  {\bibfnamefont {X.}~\bibnamefont {Wang}}, \bibinfo {author} {\bibfnamefont
  {L.}~\bibnamefont {Xiao}}, \bibinfo {author} {\bibfnamefont {H.-D.}\
  \bibnamefont {Cheng}},\ and\ \bibinfo {author} {\bibfnamefont
  {L.}~\bibnamefont {Liu}},\ }\bibfield  {title} {\bibinfo {title}
  {Optical-plus-microwave pumping in a magnetically insensitive state of cold
  atoms},\ }\href@noop {} {\bibfield  {journal} {\bibinfo  {journal} {Phy. Rev.
  A}\ }\textbf {\bibinfo {volume} {97}},\ \bibinfo {pages} {023421} (\bibinfo
  {year} {2018})}\BibitemShut {NoStop}%
\bibitem [{\citenamefont {van Zoest}\ \emph {et~al.}(2010)\citenamefont {van
  Zoest}, \citenamefont {Gaaloul}, \citenamefont {Singh}, \citenamefont
  {Ahlers}, \citenamefont {Herr}, \citenamefont {Seidel}, \citenamefont
  {Ertmer}, \citenamefont {Rasel}, \citenamefont {Eckart}, \citenamefont
  {Kajari} \emph {et~al.}}]{van2010bose}%
  \BibitemOpen
  \bibfield  {author} {\bibinfo {author} {\bibfnamefont {T.}~\bibnamefont {van
  Zoest}}, \bibinfo {author} {\bibfnamefont {N.}~\bibnamefont {Gaaloul}},
  \bibinfo {author} {\bibfnamefont {Y.}~\bibnamefont {Singh}}, \bibinfo
  {author} {\bibfnamefont {H.}~\bibnamefont {Ahlers}}, \bibinfo {author}
  {\bibfnamefont {W.}~\bibnamefont {Herr}}, \bibinfo {author} {\bibfnamefont
  {S.}~\bibnamefont {Seidel}}, \bibinfo {author} {\bibfnamefont
  {W.}~\bibnamefont {Ertmer}}, \bibinfo {author} {\bibfnamefont
  {E.}~\bibnamefont {Rasel}}, \bibinfo {author} {\bibfnamefont
  {M.}~\bibnamefont {Eckart}}, \bibinfo {author} {\bibfnamefont
  {E.}~\bibnamefont {Kajari}}, \emph {et~al.},\ }\bibfield  {title} {\bibinfo
  {title} {{Bose-Einstein condensation in microgravity}},\ }\href@noop {}
  {\bibfield  {journal} {\bibinfo  {journal} {Science}\ }\textbf {\bibinfo
  {volume} {328}},\ \bibinfo {pages} {1540} (\bibinfo {year}
  {2010})}\BibitemShut {NoStop}%
\bibitem [{\citenamefont {Liu}\ \emph {et~al.}(2018)\citenamefont {Liu},
  \citenamefont {L{\"{u}}}, \citenamefont {Chen}, \citenamefont {Li},
  \citenamefont {Qu}, \citenamefont {Wang}, \citenamefont {Li}, \citenamefont
  {Ren}, \citenamefont {Dong}, \citenamefont {Zhao} \emph
  {et~al.}}]{liu2018orbit}%
  \BibitemOpen
  \bibfield  {author} {\bibinfo {author} {\bibfnamefont {L.}~\bibnamefont
  {Liu}}, \bibinfo {author} {\bibfnamefont {D.-S.}\ \bibnamefont {L{\"{u}}}},
  \bibinfo {author} {\bibfnamefont {W.-B.}\ \bibnamefont {Chen}}, \bibinfo
  {author} {\bibfnamefont {T.}~\bibnamefont {Li}}, \bibinfo {author}
  {\bibfnamefont {Q.-Z.}\ \bibnamefont {Qu}}, \bibinfo {author} {\bibfnamefont
  {B.}~\bibnamefont {Wang}}, \bibinfo {author} {\bibfnamefont {L.}~\bibnamefont
  {Li}}, \bibinfo {author} {\bibfnamefont {W.}~\bibnamefont {Ren}}, \bibinfo
  {author} {\bibfnamefont {Z.-R.}\ \bibnamefont {Dong}}, \bibinfo {author}
  {\bibfnamefont {J.-B.}\ \bibnamefont {Zhao}}, \emph {et~al.},\ }\bibfield
  {title} {\bibinfo {title} {In-orbit operation of an atomic clock based on
  laser-cooled $^{87}$rb atoms},\ }\href@noop {} {\bibfield  {journal}
  {\bibinfo  {journal} {Nat. Commun.}\ ,\ \bibinfo {pages} {2760}} (\bibinfo
  {year} {2018})}\BibitemShut {NoStop}%
\bibitem [{\citenamefont {Aveline}\ \emph {et~al.}(2020)\citenamefont
  {Aveline}, \citenamefont {Williams}, \citenamefont {Elliott}, \citenamefont
  {Dutenhoffer}, \citenamefont {Kellogg}, \citenamefont {Kohel}, \citenamefont
  {Lay}, \citenamefont {Oudrhiri}, \citenamefont {Shotwell}, \citenamefont
  {Yu},\ and\ \citenamefont {Thompson}}]{aveline2020-Observation}%
  \BibitemOpen
  \bibfield  {author} {\bibinfo {author} {\bibfnamefont {D.~C.}\ \bibnamefont
  {Aveline}}, \bibinfo {author} {\bibfnamefont {J.~R.}\ \bibnamefont
  {Williams}}, \bibinfo {author} {\bibfnamefont {E.~R.}\ \bibnamefont
  {Elliott}}, \bibinfo {author} {\bibfnamefont {C.}~\bibnamefont
  {Dutenhoffer}}, \bibinfo {author} {\bibfnamefont {J.~R.}\ \bibnamefont
  {Kellogg}}, \bibinfo {author} {\bibfnamefont {J.~M.}\ \bibnamefont {Kohel}},
  \bibinfo {author} {\bibfnamefont {N.~E.}\ \bibnamefont {Lay}}, \bibinfo
  {author} {\bibfnamefont {K.}~\bibnamefont {Oudrhiri}}, \bibinfo {author}
  {\bibfnamefont {R.~F.}\ \bibnamefont {Shotwell}}, \bibinfo {author}
  {\bibfnamefont {N.}~\bibnamefont {Yu}},\ and\ \bibinfo {author}
  {\bibfnamefont {R.~J.}\ \bibnamefont {Thompson}},\ }\bibfield  {title}
  {\bibinfo {title} {Observation of bose-einstein condensates in an
  earth-orbiting research lab},\ }\href@noop {} {\bibfield  {journal} {\bibinfo
   {journal} {Nature}\ }\textbf {\bibinfo {volume} {582}},\ \bibinfo {pages}
  {193} (\bibinfo {year} {2020})}\BibitemShut {NoStop}%
\bibitem [{\citenamefont {Zhang}\ \emph
  {et~al.}(2009{\natexlab{a}})\citenamefont {Zhang}, \citenamefont {Cheng},
  \citenamefont {Liu},\ and\ \citenamefont {Wang}}]{zhang2009Observation}%
  \BibitemOpen
  \bibfield  {author} {\bibinfo {author} {\bibfnamefont {W.-Z.}\ \bibnamefont
  {Zhang}}, \bibinfo {author} {\bibfnamefont {H.-D.}\ \bibnamefont {Cheng}},
  \bibinfo {author} {\bibfnamefont {L.}~\bibnamefont {Liu}},\ and\ \bibinfo
  {author} {\bibfnamefont {Y.-Z.}\ \bibnamefont {Wang}},\ }\bibfield  {title}
  {\bibinfo {title} {Observation of recoil-induced resonances and
  electromagnetically induced absorption of diffuse light by cold atoms},\
  }\href@noop {} {\bibfield  {journal} {\bibinfo  {journal} {Phys. Rev. A}\
  }\textbf {\bibinfo {volume} {79}},\ \bibinfo {pages} {053804} (\bibinfo
  {year} {2009}{\natexlab{a}})}\BibitemShut {NoStop}%
\bibitem [{\citenamefont {Zhang}\ \emph
  {et~al.}(2009{\natexlab{b}})\citenamefont {Zhang}, \citenamefont {Cheng},
  \citenamefont {Xiao}, \citenamefont {Liu},\ and\ \citenamefont
  {Wang}}]{zhang2009Nonlinear}%
  \BibitemOpen
  \bibfield  {author} {\bibinfo {author} {\bibfnamefont {W.-Z.}\ \bibnamefont
  {Zhang}}, \bibinfo {author} {\bibfnamefont {H.-D.}\ \bibnamefont {Cheng}},
  \bibinfo {author} {\bibfnamefont {L.}~\bibnamefont {Xiao}}, \bibinfo {author}
  {\bibfnamefont {L.}~\bibnamefont {Liu}},\ and\ \bibinfo {author}
  {\bibfnamefont {Y.-Z.}\ \bibnamefont {Wang}},\ }\bibfield  {title} {\bibinfo
  {title} {Nonlinear spectroscopy of cold atoms in diffuse laser light},\
  }\href@noop {} {\bibfield  {journal} {\bibinfo  {journal} {Opt. Express}\
  }\textbf {\bibinfo {volume} {17}},\ \bibinfo {pages} {2892} (\bibinfo {year}
  {2009}{\natexlab{b}})}\BibitemShut {NoStop}%
\bibitem [{\citenamefont {Wang}\ \emph {et~al.}(2012)\citenamefont {Wang},
  \citenamefont {Cheng}, \citenamefont {Xiao}, \citenamefont {Zheng},
  \citenamefont {Meng}, \citenamefont {Liu},\ and\ \citenamefont
  {Wang}}]{xu2012measurement}%
  \BibitemOpen
  \bibfield  {author} {\bibinfo {author} {\bibfnamefont {X.-C.}\ \bibnamefont
  {Wang}}, \bibinfo {author} {\bibfnamefont {H.-D.}\ \bibnamefont {Cheng}},
  \bibinfo {author} {\bibfnamefont {L.}~\bibnamefont {Xiao}}, \bibinfo {author}
  {\bibfnamefont {B.-C.}\ \bibnamefont {Zheng}}, \bibinfo {author}
  {\bibfnamefont {Y.-L.}\ \bibnamefont {Meng}}, \bibinfo {author}
  {\bibfnamefont {L.}~\bibnamefont {Liu}},\ and\ \bibinfo {author}
  {\bibfnamefont {Y.-Z.}\ \bibnamefont {Wang}},\ }\bibfield  {title} {\bibinfo
  {title} {Measurement of spatial distribution of cold atoms in an integrating
  sphere},\ }\href@noop {} {\bibfield  {journal} {\bibinfo  {journal} {Chin.
  Phys. Lett.}\ }\textbf {\bibinfo {volume} {29}},\ \bibinfo {pages} {023701}
  (\bibinfo {year} {2012})}\BibitemShut {NoStop}%
\bibitem [{\citenamefont {Meng}\ \emph {et~al.}(2014)\citenamefont {Meng},
  \citenamefont {Cheng}, \citenamefont {Liu}, \citenamefont {Zheng},
  \citenamefont {Xiao}, \citenamefont {Wan}, \citenamefont {Wang},\ and\
  \citenamefont {Liu}}]{meng2014increasing}%
  \BibitemOpen
  \bibfield  {author} {\bibinfo {author} {\bibfnamefont {Y.-L.}\ \bibnamefont
  {Meng}}, \bibinfo {author} {\bibfnamefont {H.-D.}\ \bibnamefont {Cheng}},
  \bibinfo {author} {\bibfnamefont {P.}~\bibnamefont {Liu}}, \bibinfo {author}
  {\bibfnamefont {B.-C.}\ \bibnamefont {Zheng}}, \bibinfo {author}
  {\bibfnamefont {L.}~\bibnamefont {Xiao}}, \bibinfo {author} {\bibfnamefont
  {J.-Y.}\ \bibnamefont {Wan}}, \bibinfo {author} {\bibfnamefont {X.-M.}\
  \bibnamefont {Wang}},\ and\ \bibinfo {author} {\bibfnamefont
  {L.}~\bibnamefont {Liu}},\ }\bibfield  {title} {\bibinfo {title} {Increasing
  the cold atom density in an integrating spherical cavity},\ }\href@noop {}
  {\bibfield  {journal} {\bibinfo  {journal} {Phys. Lett. A}\ }\textbf
  {\bibinfo {volume} {378}},\ \bibinfo {pages} {2034} (\bibinfo {year}
  {2014})}\BibitemShut {NoStop}%
\bibitem [{\citenamefont {Zheng}\ \emph {et~al.}(2014)\citenamefont {Zheng},
  \citenamefont {Cheng}, \citenamefont {Meng}, \citenamefont {Liu},
  \citenamefont {Wang}, \citenamefont {Xiao}, \citenamefont {Wan},\ and\
  \citenamefont {Liu}}]{ZhengBC2014LargeScale}%
  \BibitemOpen
  \bibfield  {author} {\bibinfo {author} {\bibfnamefont {B.-C.}\ \bibnamefont
  {Zheng}}, \bibinfo {author} {\bibfnamefont {H.-D.}\ \bibnamefont {Cheng}},
  \bibinfo {author} {\bibfnamefont {Y.-L.}\ \bibnamefont {Meng}}, \bibinfo
  {author} {\bibfnamefont {P.}~\bibnamefont {Liu}}, \bibinfo {author}
  {\bibfnamefont {X.-M.}\ \bibnamefont {Wang}}, \bibinfo {author}
  {\bibfnamefont {L.}~\bibnamefont {Xiao}}, \bibinfo {author} {\bibfnamefont
  {J.-Y.}\ \bibnamefont {Wan}},\ and\ \bibinfo {author} {\bibfnamefont
  {L.}~\bibnamefont {Liu}},\ }\bibfield  {title} {\bibinfo {title} {A
  large-scale cold atom source in an integrating sphere},\ }\href@noop {}
  {\bibfield  {journal} {\bibinfo  {journal} {Mod. Phys. Lett. B}\ }\textbf
  {\bibinfo {volume} {28}},\ \bibinfo {pages} {1450116} (\bibinfo {year}
  {2014})}\BibitemShut {NoStop}%
\bibitem [{\citenamefont {Wan}\ \emph {et~al.}(2015)\citenamefont {Wan},
  \citenamefont {Cheng}, \citenamefont {Meng}, \citenamefont {Xiao},
  \citenamefont {Liu}, \citenamefont {Wang}, \citenamefont {Wang},\ and\
  \citenamefont {Liu}}]{wan2015Non}%
  \BibitemOpen
  \bibfield  {author} {\bibinfo {author} {\bibfnamefont {J.-Y.}\ \bibnamefont
  {Wan}}, \bibinfo {author} {\bibfnamefont {H.-D.}\ \bibnamefont {Cheng}},
  \bibinfo {author} {\bibfnamefont {Y.-L.}\ \bibnamefont {Meng}}, \bibinfo
  {author} {\bibfnamefont {L.}~\bibnamefont {Xiao}}, \bibinfo {author}
  {\bibfnamefont {P.}~\bibnamefont {Liu}}, \bibinfo {author} {\bibfnamefont
  {X.-M.}\ \bibnamefont {Wang}}, \bibinfo {author} {\bibfnamefont {Y.-N.}\
  \bibnamefont {Wang}},\ and\ \bibinfo {author} {\bibfnamefont
  {L.}~\bibnamefont {Liu}},\ }\bibfield  {title} {\bibinfo {title}
  {Non-resonant magneto-optical effects in cold atoms},\ }\href@noop {}
  {\bibfield  {journal} {\bibinfo  {journal} {Chin. Opt. Lett.}\ }\textbf
  {\bibinfo {volume} {13}},\ \bibinfo {pages} {020201} (\bibinfo {year}
  {2015})}\BibitemShut {NoStop}%
\bibitem [{\citenamefont {Wang}\ \emph {et~al.}(2015)\citenamefont {Wang},
  \citenamefont {Deng},\ and\ \citenamefont {Wang}}]{wang2015Recoil}%
  \BibitemOpen
  \bibfield  {author} {\bibinfo {author} {\bibfnamefont {W.-L.}\ \bibnamefont
  {Wang}}, \bibinfo {author} {\bibfnamefont {J.-L.}\ \bibnamefont {Deng}},\
  and\ \bibinfo {author} {\bibfnamefont {Y.-Z.}\ \bibnamefont {Wang}},\
  }\bibfield  {title} {\bibinfo {title} {Recoil-induced resonance spectroscopy
  and nondestructive temperature measurement of cold rubidium atoms inside an
  integrating sphere},\ }\href@noop {} {\bibfield  {journal} {\bibinfo
  {journal} {J. Opt. Soc. Am. B}\ }\textbf {\bibinfo {volume} {32}},\ \bibinfo
  {pages} {2441} (\bibinfo {year} {2015})}\BibitemShut {NoStop}%
\bibitem [{\citenamefont {Wang}\ \emph {et~al.}(2020)\citenamefont {Wang},
  \citenamefont {Sun}, \citenamefont {Cheng}, \citenamefont {Wan},
  \citenamefont {Meng}, \citenamefont {Xiao},\ and\ \citenamefont
  {Liu}}]{wang2020Nearly}%
  \BibitemOpen
  \bibfield  {author} {\bibinfo {author} {\bibfnamefont {X.}~\bibnamefont
  {Wang}}, \bibinfo {author} {\bibfnamefont {Y.}~\bibnamefont {Sun}}, \bibinfo
  {author} {\bibfnamefont {H.-D.}\ \bibnamefont {Cheng}}, \bibinfo {author}
  {\bibfnamefont {J.-Y.}\ \bibnamefont {Wan}}, \bibinfo {author} {\bibfnamefont
  {Y.-L.}\ \bibnamefont {Meng}}, \bibinfo {author} {\bibfnamefont
  {L.}~\bibnamefont {Xiao}},\ and\ \bibinfo {author} {\bibfnamefont
  {L.}~\bibnamefont {Liu}},\ }\bibfield  {title} {\bibinfo {title} {Nearly
  nondestructive thermometry of labeled cold atoms and application to isotropic
  laser cooling},\ }\href@noop {} {\bibfield  {journal} {\bibinfo  {journal}
  {Phys. Rev. Applied}\ }\textbf {\bibinfo {volume} {14}},\ \bibinfo {pages}
  {024030} (\bibinfo {year} {2020})}\BibitemShut {NoStop}%
\end{thebibliography}%

\end{document}